\documentclass[a4paper,12pt]{article}

\def\displayandname#1{\rlap{$\displaystyle\csname #1\endcsname$}%
                      \qquad \texttt{\char92 #1}}

\usepackage{epsfig}

\begin{document}

\title{Renormalization and dressing
 in quantum field theory}

\author{Eugene V. Stefanovich \and \small \emph{2255 Showers Dr., Apt. 153, 
Mountain View, CA 94040, USA} \and \small $eugene\_stefanovich@usa.net$}

\maketitle

\begin{abstract}
We illustrate the mass and charge renormalization procedures in quantum
field theory using, as an example, a simple model of interacting
electrons and photons. It is shown how addition of infinite
renormalization counterterms to the Hamiltonian helps to obtain finite
and accurate results for the $S$-matrix. In order to remove the
ultraviolet divergences from the Hamiltonian, we apply 
the Greenberg-Schweber ``dressing transformation'' 
and
the  G\l azek-Wilson ``similarity renormalization''. 
The resulting ``dressed particle''
Hamiltonian is finite 
in all orders of the perturbation theory and yields accurate $S$-matrix
and bound state energies. The bare and virtual particles are removed
from the theory, and physical dressed particles interact via direct
action-at-a-distance.
\end{abstract}

\section{ Introduction}

Consistent unification of relativity and quantum mechanics remains an
unsolved theoretical problem in spite of many efforts applied to its
solution in the 20th century. 
The fundamental difference between relativistic and non-relativistic
physics follows from the famous Einstein's formula $E = mc^2$. This
formula, in particular, implies that if a
system of particles has sufficient energy $E$ of their relative
motion, then this energy may be converted to the mass $m$ of newly
created particles. 
Generally, there is no limit on how many particles can be created 
in collisions, so any
realistic quantum mechanical description of high-energy systems
should involve states with any number of particles from zero to
infinity.
The number of particles is not conserved during time evolution. The
most familiar example of such a behavior is the emission and absorption
of light (photons) in electrodynamics.

First attempts to describe relativistic quantum systems were
undertaken immediately after creation of the
formalism of quantum mechanics in 1920's. A quantum theory of the
electromagnetic field was constructed by
quantization of the classical Maxwell electrodynamics. In lowest
perturbation orders, this theory agreed well with experimental
results. 
However, perturbative calculations for the $S$-matrix  did not work in
higher orders, in particular, due to ultraviolet divergences.

The way to calculate  the $S$-matrix in QFT
accurately in all orders was provided by the renormalized QED formulated by Feynman, Schwinger,
and Tomonaga in the late 1940's. However this approach created a host of
other problems.  According
to the prevailing interpretation, the creation and annihilation
 operators present in
the Hamiltonian and Lagrangian of QED correspond to  \emph{bare}
particles having infinite
masses and charges.\footnote{Infinite quantities appear frequently in
the renormalization theory. In order to avoid them in practical
calculations,
one often uses \emph{regularization}, e.g., cutting off all 
integrals at large integration momenta. In a regularized theory
 all quantities become
finite. However, the regularization is an artificial trick, and in
order to get rigorous and accurate results one should take the infinite
limit of the cutoff momentum.
As will be discussed later in the paper,
the goal of renormalization is to introduce certain cancellations
between (finite) regularized quantities, so that when the integration
cutoff is lifted the physically relevant parameters have well
defined finite limits. Without indicating this explicitly in calculations, we will
always assume that the regularization and the taking the cutoff to
infinity steps were properly executed. A quantity will be called
finite (infinite) if it has (does not have)  a finite value in the limit of
infinite cutoff.}    However, the bare
particles  have never been directly observed in
experiments. They are believed to be surrounded by clouds of
virtual photons and electron-positron pairs, thus forming  complex
objects  called \emph{dressed} particles. The dressed particles are
supposed to be the eigenstates of the full Hamiltonian. They have
finite experimentally observable masses and charges.  The problem is that the bare particle Hamiltonian 
 of QFT 
is formally infinite. Although, these infinities cancel out when the 
$S$-matrix is calculated, the Hamiltonian is useless if one wants to
calculate the time evolution or to find wavefunctions of bound states
via diagonalization procedure. 

Two lines of research were initiated to cope with these problems.
The \emph{dressed particle} approach was suggested by
Greenberg and Schweber \cite{GS}. Their goal was to get rid of the bare particles
and to express the entire formalism of QFT through observable dressed
particles only. The operators for dressed particles were sought as
unitary transforms of bare particle operators. 
From another direction, 
G\l azek and Wilson \cite{Glazek} introduced the \emph{similarity renormalization}
formalism.
The idea was to apply a unitary transformation to the Hamiltonian
in order to ensure that
interaction potentials rapidly fall off as functions of energy differences, and
to guarantee that all loop integrals are convergent.
The similarity and dressing transformations are strikingly
similar.
They can be combined 
into one unitary \emph{similarity-dressing} transformation which achieves two goals at once: the
theory can be expressed in terms of real dressed particles and the
Hamiltonian can be made finite in all perturbation orders.
At the same time, the accurate and well-tested $S$-matrix of the
renormalized theory remains intact. This combined approach was
developed in refs. \cite{infinities, Stefanovich_Mink, mybook} and dubbed the relativistic quantum dynamics, or RQD.

In this paper we will discuss the reasons why the renormalization
difficulties (e.g., the absence of a well-defined Hamiltonian and
unsatisfactory treatment of the time evolution) persist in current
relativistic quantum field theories. To avoid
unnecessary mathematical complications, we will be working with a
simple model theory which, nevertheless, shares some important
features with QED. In section 9 we will explain how the
similarity-dressing transformation of this model theory leads to a
well defined (perturbatively) finite Hamiltonian that can be used for all kinds of
quantum mechanical calculations ($S$-matrix, bound states, time
evolution, etc.) without the need for renormalization. In addition,
the RQD formalism does not use the dubious notions of bare and virtual
particles.

\section{ Bound states, time evolution, and scattering}
\label{sc:bound-states}

In this paper we will be concerned with   three types
of phenomena that are normally studied in physical
 experiments: bound states, time evolution, and scattering.
These three  areas account for the most of
 experimental information available about fundamental particles and
their interactions.
 The key theoretical quantity involved 
in quantum mechanical description of these phenomena is the 
Hamilton operator $H$. 
The energies $E_n$ and state vectors $| \Psi \rangle_n$ 
of bound states  can be found as eigenvalues and eigenvectors of the Hamiltonian

\begin{eqnarray}
   H |\Psi \rangle_n = E_n | \Psi \rangle_n
\label{eq:1}
\end{eqnarray}

The development of the state vector $| \Psi
\rangle $ from time $t'$ to time $t$ is described by the \emph{time
evolution operator} $\exp(iHt)$

\begin{eqnarray}
|\Psi (t) \rangle = e^{iH(t-t')}| \Psi (t') \rangle
\label{eq:8.60}
\end{eqnarray}

\noindent If the eigenvalues $E_n$ and eigenvectors $| \Psi
\rangle_n$  of the Hamiltonian are known and
the initial state is represented as a sum (and/or integral) over the
basis states

\begin{eqnarray}
| \Psi (t') \rangle = \sum_n C_n(t') | \Psi \rangle_n
\end{eqnarray}

\noindent then the time evolution can be calculated as

\begin{eqnarray}
|\Psi (t) \rangle 
&=& \sum_n C_n(t') e^{iE_n (t-t')} | \Psi \rangle_n
\label{eq:8.61a}
\end{eqnarray}

\noindent Unfortunately, in most cases, the full spectrum of the
Hamiltonian is not known,
and the time evolution is difficult to predict.

This difficulty, however, is not that disappointing, because
experimental observations of the time evolution in subatomic world
are even more difficult than calculations.
Most experiments in high energy physics are performed by
preparing free particles or their bound states (like hydrogen atoms or
deuterons), bringing them 
into collision, and studying the properties of free particles or bound states leaving
the region of collision. In
these
experiments, it is not possible to
observe the time evolution during interaction: particle reactions
occur almost instantaneously and one can only register the  reactants and
products which move freely before and after the
collision. This gives a lucky break for theoreticians:  
In such situations
the theory is not required to describe the detailed dynamics of
particles during the short interval of collision. It is sufficient
to provide a mapping of free states before interaction onto the
free states after the interaction. 
To describe  scattering experiments, one needs only the formula for the time evolution 
from the remote past
$t' \ll 0$ to the distant future
$t \gg 0$ 

\begin{eqnarray}
 e^{iH (t - t')}
&=&  e^{iH_0t}
S e^{-iH_0t'}
\label{eq:8.65}
\end{eqnarray}

\noindent where the $S$\emph{-operator} \label{s-oper} 
 is
defined by 

\begin{eqnarray}
S 
  &=& \lim_{t \to \infty} \lim_{t' \to - \infty} e^{-iH_0 t} 
e^{iH(t-t')}
e^{iH_0t'}
\label{eq:s-oper}
\end{eqnarray}

\noindent One can read the right hand side of eq. (\ref{eq:s-oper}) from right to
left as a sequence of three steps: (i) the 
non-interaction evolution of the system from time $t'$
in the past to 0; (ii) the sudden jump at $t=0$ described by the
$S$-operator;
(iii) the free evolution from $t=0$ to the future time $t$.

 The most effective technique available for 
calculations of the $S$-operator is the
perturbation theory which can be written in
many equivalent forms.\footnote{In this paper we are not discussing the
complicated issue of the convergence of perturbative expansions. We
will assume that all perturbative series do converge.} The Dyson time-ordered expansion provides the
most economical expressions that can be encoded in familiar Feynman
diagrams.
However, for the discussion in this paper, we found more useful
two other perturbative expressions  which differ from the Dyson's
formula only by re-shuffling the terms. The ``old-fashioned'' 
 formula for the $S$-operator is

\begin{eqnarray}
    S &=& 1 + i\int_{-\infty}^{+\infty} V(t)\,dt
          -  \int_{-\infty}^{+\infty} V(t)\,dt \int_{-\infty}^{t} V(t')\,dt'
          + \ldots
\label{eq:8.68}  
\end{eqnarray}

\noindent  where $V$ is the interaction part of the total Hamiltonian $H = H_0 +V$,

\begin{eqnarray}
V(t) = e^{-iH_0t} V e^{iH_0t} e^{-\epsilon
|t|}
\label{eq:8.67}
\end{eqnarray}

\noindent and the factor $e^{-\epsilon
|t|}$ in the limit $\epsilon \to 0$ serves for adiabatic switching the interaction on and
off. Operators with $t$-dependence determined by the free Hamiltonian
$H_0$ as in eq. (\ref{eq:8.67}) 
 will be called
\emph{regular}. 
Using convenient symbols for $t$-integrals 
\label{underline}

\begin{eqnarray*}
 \underline{Y(t)} &\equiv&
\int_{-\infty }^{t} Y(t') d t' \\
 \underbrace{Y(t)}&\equiv&
\int_{-\infty }^{+\infty} Y(t') d t'
\end{eqnarray*}

\noindent  formula (\ref{eq:8.68}) can be written compactly as

\begin{eqnarray}
S   =  1 +
\underbrace{\Sigma(t)}
\label{eq:8.69}
\end{eqnarray}

\noindent  where

\begin{eqnarray}
    \Sigma(t) &=&
            i V(t)
          -  V(t) \underline{V(t)}
          - i V(t) \underline{V(t) \underline{V(t)}}
          +  V(t) \underline{V(t) \underline{V(t) \underline{V(t)}}}
          +   \ldots
\label{eq:8.70} 
\end{eqnarray}

Another equivalent
expression for $S$ was suggested by  Magnus \cite{Magnus}

\begin{eqnarray}
   S = e^{i\underbrace{F(t)}}
\label{eq:8.71}
\end{eqnarray}

\noindent Here the
  Hermitian operator $F(t)$ can be represented as a
series of multiple commutators with $t$-integrations

\begin{eqnarray}
   F(t)  &=&  V(t)
   - \frac{i}{2 }[\underline{V(t)},V(t)]
   -\frac{1}{6}[\underline{\underline{V(t)},[V(t)},V(t)]]
   -\frac{1}{6}[\underline{[\underline{V(t)},V(t)]},V(t)]
\nonumber \\
 &\mbox{ }&  +\frac{i}{12}[\underline{\underline{\underline{V(t)},[[V(t)},V(t)]},V(t)]]
\nonumber \\
   &+&\frac{i}{12}[\underline{[\underline{\underline{V(t)},[V(t)},V(t)]]},V(t)]
   +\frac{i}{12}[\underline{\underline{[\underline{V(t)},V(t)]},[V(t)},V(t)]]
   +\ldots
\label{eq:8.72}
\end{eqnarray}

\noindent One important advantage of this representation is that the
$S$-operator (\ref{eq:8.71})
is manifestly unitary.
It follows from equations (\ref{eq:8.69}) and (\ref{eq:8.71})
 that operators $\Sigma(t)$ and $F(t)$ are
related to each other

\begin{eqnarray}
    F(t) &=& -i \frac{d}{dt} \log (1 + \underline{\Sigma(t)})
\label{eq:8.74}
\end{eqnarray}

\noindent so finding $F(t)$ or $\Sigma(t)$ are equivalent tasks.

The $S$-operator and the Hamiltonian provide two different ways to
describe dynamics.
The $S$-operator represents only ``integrated''
time evolution from the remote past to the distant future. 
The knowledge of the $S$-operator is sufficient to
calculate the scattering cross-sections as well as energies and
lifetimes of stable and metastable bound states.\footnote{The latter two
quantities are represented by positions of poles of the $S$-operator
on the complex energy plane.} However, in order to describe the time
evolution and the wavefunctions of bound states 
the full interacting Hamiltonian $H$
is required.

It can be shown \cite{Ekstein} that two Hamiltonians $ H $ and 
$ H' $  related to each other by a unitary transformation 
$e^{i\Phi}$

\begin{eqnarray*}
H' &=& e^{i\Phi} H e^{-i\Phi} 
\end{eqnarray*}

\noindent yield
the same scattering $S' = S$  as long as condition

\begin{eqnarray}
\lim _{t \to  \pm \infty}  e^{-iH_0 t} \Phi e^{iH_0t}= 0
\label{eq:8.75}
\end{eqnarray}

\noindent is satisfied.  Such Hamiltonians $H$ and $H'$ are called
\emph{scattering-equivalent}.
The energy spectra of two scattering equivalent Hamiltonians
are identical.
However, the eigenvectors are different and, according to eq. (\ref{eq:8.61a}),  the
corresponding descriptions of
dynamics  are different as well. Therefore scattering-equivalent
theories may be not
\emph{physically} equivalent.

Calculations of bound states, time evolution and scattering is a
routine practice in non-relativistic quantum mechanics. 
However, the situation is less certain 
in relativistic quantum field theories. As mentioned in Introduction, the
Hamiltonian of renormalized QFT is infinite. Therefore,
it is not immediately clear if the above formulas (\ref{eq:1}) - (\ref{eq:8.74}) 
remain valid for the high energy
relativistic phenomena, and what modifications, if any, should be
introduced in quantum theory to take into account the variable
number 
of particles. 
In this paper we illustrate the difficulties encountered in quantum
field theories by analyzing  a simple   model
 theory with variable number of particles. We will demonstrate that
 the  Hamiltonian can be redefined so that there is no
need for renormalization and  usual
quantum mechanical techniques remain applicable even in the relativistic
case.

\section { Model theory}
\label{sc:model-theory}

Our model  theory   describes
two kinds of particles. These are
massive spinless fermions which will be called \emph{electrons} and
 massless bosons with zero
helicity, which  will be called \emph{photons}. Here we disregard the
spin and polarization degrees of freedom as they are not so important
for the discussion of renormalization.  
To allow for creation and annihilation of particles, the system is
 described in the Fock space which is built 
as a direct sum of sectors with various numbers of particles.
For example, if we denote $|0 \rangle$ the no-particle vacuum state, 
$\mathcal{H}_{el}$
the one-electron
Hilbert space and
$\mathcal{H}_{ph}$ the one-photon Hilbert space, 
then 
the Fock space can be written as an infinite direct sum

\begin{eqnarray}
\mathcal{H}  &=& | 0 \rangle \oplus \mathcal{H}_{el} 
\oplus  \mathcal{H}_{ph} 
\oplus  (\mathcal{H}_{el} \otimes \mathcal{H}_{ph}) 
\oplus (\mathcal{H}_{el} \otimes_{asym} \mathcal{H}_{el})
\oplus (\mathcal{H}_{ph} \otimes_{sym} \mathcal{H}_{ph})
\ldots 
\end{eqnarray}

The
anticommutation and commutation relations for particle creation and
annihilation operators ($a^{\dag}_{\mathbf{p}}$, $a_{\mathbf{p}}$ for
electrons and $c^{\dag}_{\mathbf{p}}$, $c_{\mathbf{p}}$ for photons,
respectively)
are, as usual,

\begin{eqnarray}
\{a_{\mathbf{p}},a^{\dag}_{\mathbf{p}'}\} &=& 
\delta (\mathbf{p}-\mathbf{p}')
\label{eq:aa}
\end{eqnarray}
\begin{eqnarray}
[c_{\mathbf{k}},c^{\dag}_{\mathbf{k}'}]
&=&  \delta (\mathbf{p}-\mathbf{p}') 
\label{eq:cc}\\
  \{a_{\mathbf{p}},a_{\mathbf{p}'}\}
&=& \{a^{\dag}_{\mathbf{p}},a^{\dag}_{\mathbf{p}'}\} = 0 
\label{eq:aa2}
\end{eqnarray}
\begin{eqnarray}
 [c_{\mathbf{k}},c_{\mathbf{k}'}]
= [c^{\dag}_{\mathbf{k}},c^{\dag}_{\mathbf{k}'}] = 0
\label{eq:cc2}
\end{eqnarray}
\begin{eqnarray*}
  [a^{\dag}_{\mathbf{p}},c^{\dag}_{\mathbf{k}}]&=&
[a^{\dag}_{\mathbf{p}},c_{\mathbf{k}}] =
[a_{\mathbf{p}},c^{\dag}_{\mathbf{k}}] =
[a_{\mathbf{p}},c_{\mathbf{k}}] = 0.
\end{eqnarray*}

\noindent The full
Hamiltonian  $ H  = H_0 + V_1 $ is the sum of  the free
Hamiltonian 

\begin{eqnarray}
    H_0 &=&  \int  d\mathbf{p} \omega_{\mathbf{p} }
  a^{\dag}_{\mathbf{p}}a_{\mathbf{p}}
+  \int \limits_{\mathbf{k} \neq 0} d\mathbf{k} |\mathbf{k}|   c^{\dag}_{\mathbf{k}}
c_{\mathbf{k}}
\label{eq:free-h0}
\end{eqnarray}

\noindent (where $\omega_{\mathbf{p}}= \sqrt{\mathbf{p}^2 + m^2}$ and $|\mathbf{k}|$
are one-particle energies of electrons and photons, respectively)  
and the interaction, which we choose in the following 
form 

\begin{eqnarray}
    V_1 &=&  e (2\pi)^{-3/2} \int \limits_{\mathbf{k} \neq 0}
     \frac{d\mathbf{p}d\mathbf{k}}{\sqrt{|\mathbf{k}|}}
a^{\dag}_{\mathbf{p}}c^{\dag}_{\mathbf{k}}a_{\mathbf{p+k}}
+ e  (2\pi)^{-3/2} \int \limits_{\mathbf{k} \neq 0}
    \frac{d\mathbf{p}d\mathbf{k}}{\sqrt{|\mathbf{k}|}}
a^{\dag}_{\mathbf{p}}a_{\mathbf{p-k}}
     c_{\mathbf{k}}
\label{eq:9.69}
\end{eqnarray}

\noindent The coupling constant $e$ is the absolute value of the electron
charge. 
Here and in what follows the \emph{perturbation order} \label{pert-order}
of an operator (= the power of the coupling constant $e$) 
is shown by the subscript. For example, the free Hamiltonian
$H_0$ does not depend on $e$, so it is of zero perturbation order; $V_1$
is of the  first perturbation order, etc.
The number of electrons is conserved  by the interaction (\ref{eq:9.69}),
but the number of photons is not conserved. So, this
theory is capable of describing important processes of the  emission
and absorption of photons. 

In this paper we will discuss ultraviolet divergences associated
with the interaction (\ref{eq:9.69}). However, we will skip completely the
discussion of ``infrared
divergences''
which are related to the zero mass of photons and singularities
$|\mathbf{k}|^{-1/2}$ in (\ref{eq:9.69}).  The easiest way to avoid infrared problems in practical
calculations is to assign a small non-zero mass to photons.

In section \ref{ss:e-e-scatt} we are going  to calculate the
scattering operator 
(\ref{eq:8.69}) with the
above Hamiltonian. Before doing that, some remarks are in order.   
The
operator $S$ is obtained as a sum of products (\ref{eq:8.70}) (or
commutators (\ref{eq:8.72})) of
interactions $V_1(t) =
e^{-iH_0t} V_1 e^{iH_0t}$  with $t$-integrations.   Each term in these
expressions can be written as a
normally ordered product of $N$ 
creation operators $\alpha^{\dag}$ and $M$ annihilation operators
$\alpha$.\footnote{Here symbols $\alpha^{\dag}$ and 
$\alpha$ refer to generic creation and annihilation operators without
specifying the type of the particle. The pair of integers $(N,M)$ will be referred to as the
\emph {index} \label{index} of the term $V_{NM}$. }

\begin{eqnarray}
 V_{NM}(t)     &=&
 \int   [d\mathbf{q}]
   D_{NM}[\mathbf{q}]  e^{itE_{NM}[\mathbf{q}]}
    \delta
(\mathbf{P}_{NM}[\mathbf{q}])  
\alpha^{\dag}_{\mathbf{q}'_1} \ldots
\alpha^{\dag}_{\mathbf{q}'_N}
\alpha_{
\mathbf{q}_1} \ldots \alpha_{\mathbf{q}_M}
\label{eq:9.49}
\end{eqnarray}

\noindent where 
integration is carried  over  momenta of all created and annihilated
particles $[d\mathbf{q}]$. 
  The momentum conservation law is guaranteed by the
delta function in (\ref{eq:9.49}) whose argument is
the sum of momenta of
created particles minus the sum of momenta of annihilated
particles

\begin{eqnarray}
 \mathbf{P}_{NM}(\mathbf{q}'_1, \ldots, \mathbf{q}'_N, \mathbf{q}_1,  \ldots,
\mathbf{q}_M)
 \equiv \sum_{i=1}^N \mathbf{q}'_i
 - \sum_{j=1}^M \mathbf{q}_j
\label{eq:9.51a}
\end{eqnarray}

\noindent Usually, we will perform explicit integration over one momentum
$\mathbf{q}'_1$
which  removes the delta function and expresses $\mathbf{q}'_1$ as a
linear function of other momenta.
The argument of the exponent in (\ref{eq:9.49}) contains the \emph{energy function} 

\begin{eqnarray}
 E_{NM}(\mathbf{q}'_1, \ldots, \mathbf{q}'_N, \mathbf{q}_1,  \ldots,
\mathbf{q}_M)
 \equiv \sum_{i=1}^N \omega_{\mathbf{q}'_i}
 - \sum_{j=1}^M \omega_{\mathbf{q}_j}
\label{eq:9.51}
\end{eqnarray}

\noindent which is 
the difference of energies of particles created and
annihilated by $V_{NM}$.  $D_{NM}$ is a numerical
\emph{coefficient function}.

 Suppose that a  term $V_{NM}$ has coefficient function 
$D_{NM}$, then we introduce a useful notation \label{circ} $V_{NM} \circ
\zeta$ for the
operator whose coefficient function $D_{NM}'$ is a product of $D_{NM}$ and 
a function $\zeta$ of the same arguments

 \begin{eqnarray*}
D_{NM}'[\mathbf{q}] &=& D_{NM}[\mathbf{q}] \zeta [\mathbf{q}]
\end{eqnarray*}

\noindent Then, 
a $t$-dependent  regular term $V_{NM}(t)$  can be written as

\begin{eqnarray*}
V_{NM}(t) &=& e^{-iH_0t} V_{NM} e^{iH_0t} \\
&=& V_{NM} \circ e^{-iE_{NM}t}
\end{eqnarray*}

\noindent and its definite $t$-integral is

\begin{eqnarray}
 \underbrace{V_{NM}(t)} &\equiv& \int _{-\infty}^{\infty} V_{NM}(t) dt
\nonumber\\
&= &  2 \pi  V_{NM} \circ \delta(E_{NM})
\label{eq:9.52}
\end{eqnarray}

\noindent Eq. (\ref{eq:9.52}) means that each term in $ \underbrace{V_{NM}(t)}$ 
is non-zero only on the hypersurface
 of solutions of the equation

\begin{eqnarray*}
 E_{NM}(\mathbf{q}'_1, \ldots, \mathbf{q}'_N, \mathbf{q}_1,  \ldots,
\mathbf{q}_M) =  0
\end{eqnarray*}

\noindent if such solutions exist. This hypersurface in the  momentum space  
is called the \emph{energy shell} \label{energy-shell} of the
term $V_{NM}(t)$.  Note that the scattering operator (\ref{eq:8.69})
 is non-trivial only on the energy
shell, i.e., where the energy conservation condition holds.

\section {Three types of operators in the Fock space.}
\label{ss:three-types}

In this section we would like to get a further insight into the nature of 
operators in the Fock space by dividing them  into three groups
depending 
on their index $(N,M)$. 
We will call these types of operators   \emph{renorm},   \emph{phys},
and 
\emph{unphys}.

\bigskip

\textbf{Renorm operators} \label{oper-renorm}  have either index (0,0) 
(a numerical constant $C$) or index (1,1) in which case the same type
of particle is
created and annihilated.  The most general form of a renorm operator in
our theory is\footnote{Here we write just the operator
structure of $R$ omitting all numerical factors, indices, 
integration and summation signs. Note also that the interaction
operator (\ref{eq:9.69}) commutes with the operator of charge $Q =
-e \int d \mathbf{p} a^{\dag}_{\mathbf{p}}a_{\mathbf{p}}$, so any
product or commutator of terms derived from $V_1$ should also commute
with $Q$. This is not true for terms like $a^{\dag}c$ and $c^{\dag}a$,
so they are not allowed in the theory.}

\begin{eqnarray}
R &=&
a^{\dag} a +  c^{\dag} c +C
\label{eq:9.53}
\end{eqnarray}

\noindent The free Hamiltonian (\ref{eq:free-h0}) is an
example of a renorm operator.
The class of renorm operators is characterized by the property that
the energy function (\ref{eq:9.51}) is identically zero. 
 So, renorm operators always have an non-empty energy shell where they do
not vanish. Regular renorm operators do not depend on $t$.

\bigskip

\textbf{Phys  operators} \label{oper-phys} have at least two creation operators and
at least two destruction operators (index $(N,M)$ with $N \geq 2 $
and $ M \geq 2$). In  this case the energy shell is non-empty. 
For example, the energy shell for  phys  operator
$
a^{\dag}_{\mathbf{p} + \mathbf{k}} a^{\dag}_{\mathbf{q} -
\mathbf{k}} 
a_{\mathbf{p}}
a_{\mathbf{q}}
$
  is given by the set of solutions of equation
$
\omega_{\mathbf{p} + \mathbf{k}} + \omega_{\mathbf{q} - \mathbf{k}} 
= \omega_{\mathbf{p}}
+ \omega_{\mathbf{q}}
$ which is not empty.

\bigskip

\textbf{Unphys operators} \label{oper-unph}  have index $(1,N \geq 2)$ or
$(N \geq 2,1)$.
 It can be
shown that the energy shell is empty for unphys operators. For
example, the interactions (\ref{eq:9.69}) are unphys. The energy shell equation
for these
 two terms $
 \omega_{ \mathbf{p}} + |\mathbf{k}| =  \omega_{ \mathbf{p+k}} 
$
does not have a solution, because there are no photon states with zero
momentum.

  Renorm, phys, and unphys operators  exhaust all possibilities in our
theory, 
therefore
any regular operator $V$ must have a unique decomposition 

\begin{eqnarray*}
 V(t) =  V^{ren} + V^{unp}(t) + V^{ph}(t)
\end{eqnarray*}

\noindent The rules for calculations of commutators, derivatives and
$t$-integrals with different operator types  are
summarized in  Table 9.2. For example, the $t$-integrals
of 
phys and unphys
operators are given by formula

\begin{eqnarray}
\underline{V(t)} &=& V(t) \circ \frac{i }{E_{V}}
\label{eq:9.56}   
\end{eqnarray}

\begin{table}[h]
\caption{Operations with regular operators in the Fock space. 
(Notation: P=phys, U=unphys,  R=renorm,
NR=non-regular.)}
\begin{tabular*}{\textwidth}{@{\extracolsep{\fill}}ccccccc}
\hline
 Type of operator            &  & &  & & & \cr
 $A$  & $[A,P] $  &  $[A,U]$    &   $[A,R]$ &   $\frac{dA}{dt}$ &
$\underline{A}$ & $\underbrace{A}$ \cr
                  &       &         &        &          &         &  \cr
\hline
 P                & P     & P+U   &  P    & P     &    P    & P \cr
 U                & P+U & P+U+R &  U  &  U        &
U    & 0 \cr
 R                & P     &  U      &  R   &  0        &    NR   &
$\infty$ \cr
\hline
\end{tabular*}
\end{table}

\section { Diagrams in the model theory.}
\label{ss:draw-diagram}

Our goal in this section is to introduce the diagram technique
which greatly facilitates 
perturbative calculations of scattering operators (\ref{eq:8.70}) and (\ref{eq:8.72}).
Let us graphically represent each term in the interaction operator (\ref{eq:9.69})
as
a \emph{vertex} (see Fig. 1). \label{diagram}
Each particle operator in (\ref{eq:9.69}) is represented as
an oriented \emph{line} \label{line} or arrow. The line corresponding
to the annihilation operator  enters the vertex, and the line
corresponding to the  creation
operator  leaves the vertex. 
Electron lines are shown by full arcs 
 and photon lines are shown by broken arrows. Each line is marked
by the momentum label of the corresponding particle operator.
 Free ends of the electron lines
are attached to  the vertical  ``order bar'' on the left hand
side of the diagram.  The order of these \emph{external lines}
\label{extern} (from bottom to top of the
diagram) corresponds to the order of particle operators in the
 interaction term  (from right to left).    An additional numerical factor  is
indicated in the upper
left corner of the diagram.

\begin{figure}
\epsfig {file=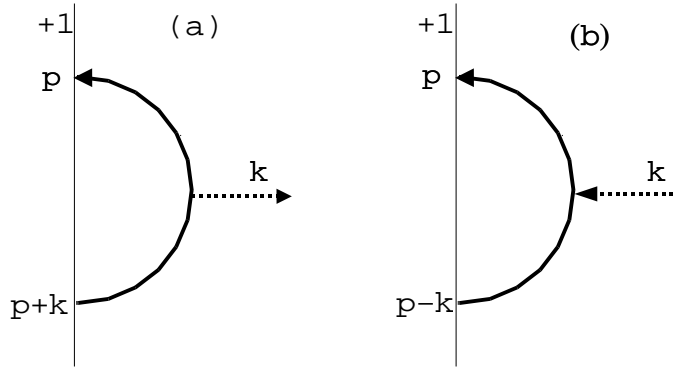}
\caption{Diagram representation of two terms in the interaction
operator $V_1$ (eq. (\ref{eq:9.69})).}
\end{figure}

The $t$-integral  $\underline{V_1(t)}$
differs from $V_1(t)$ only by the factor
$i  E_{V_1}^{-1}$ (see eq. (\ref{eq:9.56})) which is represented in the diagram
 by drawing a box that crosses all
external lines. A line entering (leaving) the box
contributes its energy with the negative (positive) sign to the energy
function $E_{V_1}$.
The diagram representation of the integral 

\begin{eqnarray}
\underline{V_1(t)} &=& \frac{ie }{(2\pi)^{3/2}} \int
    \frac{d\mathbf{p}d\mathbf{k}}{\sqrt{|\mathbf{k}|}}
\frac{e^{-it(\omega_{ \mathbf{p}} + |\mathbf{k}| 
-  \omega_{ \mathbf{p+k}})}}{\omega_{ \mathbf{p}} + |\mathbf{k}| 
-  \omega_{ \mathbf{p+k}}} 
a^{\dag}_{\mathbf{p}}c^{\dag}_{\mathbf{k}}a_{\mathbf{p+k}}
\nonumber\\
&+& \frac{ie  }{(2\pi)^{3/2}} \int
     \frac{d\mathbf{p}d\mathbf{k}}{\sqrt{|\mathbf{k}|}}
\frac{e^{-it(\omega_{ \mathbf{p}} - |\mathbf{k}| 
-  \omega_{ \mathbf{p-k}})}}{\omega_{ \mathbf{p}} - |\mathbf{k}| 
-  \omega_{ \mathbf{p-k}}} 
a^{\dag}_{\mathbf{p}}a_{\mathbf{p-k}}
     c_{\mathbf{k}}
\label{eq:9.70}
\end{eqnarray}

\noindent is shown in Fig. 2.

\begin{figure}
\epsfig {file=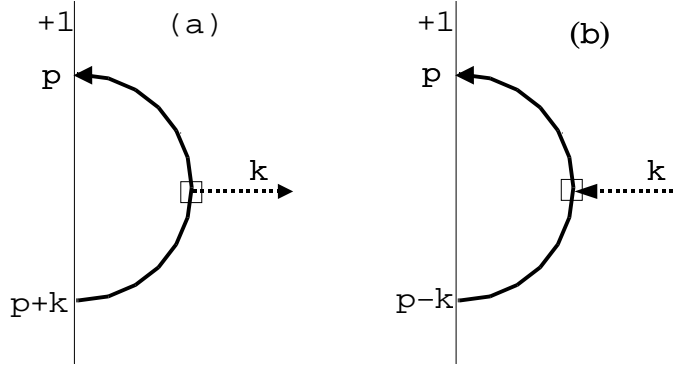}
\caption{$t$-integral $\underline{V_1(t)}$}
\end{figure}

The product of two operators $AB$ is represented 
  by simply placing diagram $B$ below diagram $A$ and attaching the
external electron lines of both diagrams to the same order bar. 
For example, the diagram for the product of
the second term in (\ref{eq:9.69}) (Fig. 1(b)) and the first term 
in (\ref{eq:9.70}) (Fig. 2(a))

\begin{eqnarray}
V_1 \underline{V_1} &\propto& 
(a^{\dag}_{\mathbf{p}}a_{\mathbf{p-k}} c_{\mathbf{k}})
(a^{\dag}_{\mathbf{q}} c^{\dag}_{\mathbf{k}'} a_{\mathbf{q+k'}}  ) +
\ldots
\label{eq:9.71}
\end{eqnarray}

\noindent  is shown in Fig. 3(a).  This product should be
further converted to the normal form, i.e., all incoming
 lines should be positioned below the outgoing lines.
Due to the  relations
(\ref{eq:aa}) - (\ref{eq:cc2}) each exchange of positions of the electron particle
operators (full external lines on the diagram) changes the total
sign of the expression. Each permutation of annihilation
and creation operators (incoming and outgoing lines) of similar
particles
 creates an additional  expression
and a new diagram in which the swapped lines are joined together.
Using these rules we first move the photon operators in (\ref{eq:9.71}) to the
rightmost positions, move the operator
$a^{\dag}_{\mathbf{q}}$ to the leftmost position, and add another
term due to the anticommutator
$\{a_{\mathbf{p-k}}, a^{\dag}_{\mathbf{q}} \} = 
\delta (\mathbf{q - p+k})$.

\begin{figure}
\epsfig {file=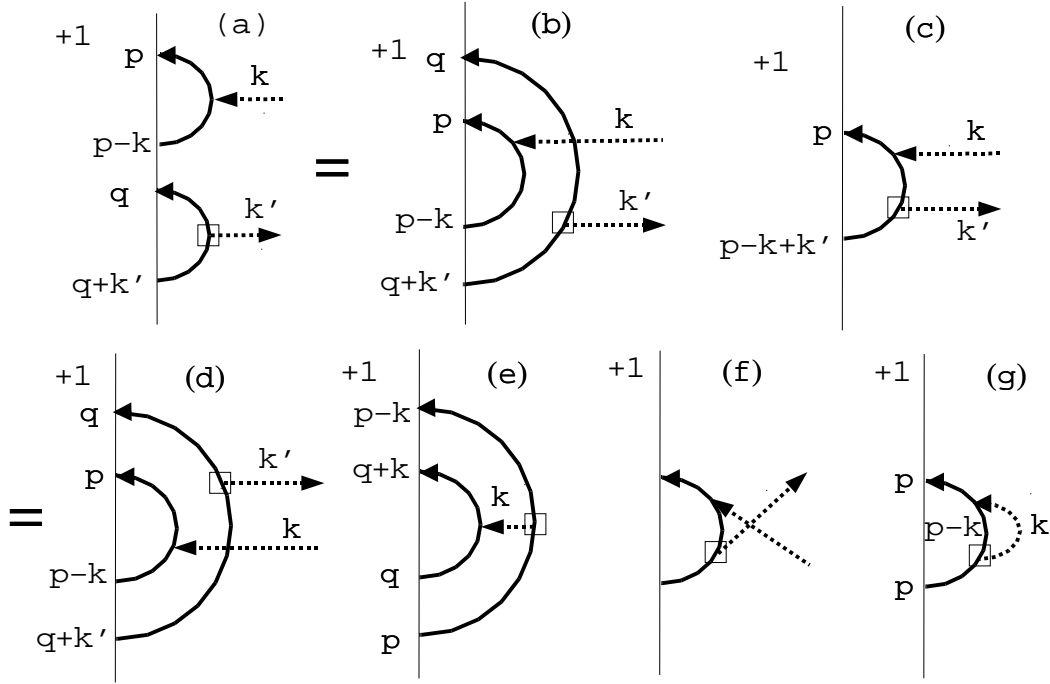}
\caption{Normal product of operators in Fig. 1(b) and 2(a).}
\end{figure} 

\begin{eqnarray}
V_1 \underline{V_1} &\propto& 
 a^{\dag}_{\mathbf{q}} a^{\dag}_{\mathbf{p}}a_{\mathbf{p-k}} 
  a_{\mathbf{q+k'}} c_{\mathbf{k}}  c^{\dag}_{\mathbf{k}'}
+ \delta(\mathbf{q} - \mathbf{p+k})a^{\dag}_{\mathbf{p}} 
  a_{\mathbf{q+k'}} c_{\mathbf{k}}  c^{\dag}_{\mathbf{k}'} \nonumber\\
&=& a^{\dag}_{\mathbf{q}} a^{\dag}_{\mathbf{p}}a_{\mathbf{p-k}} 
  a_{\mathbf{q+k'}} c_{\mathbf{k}}  c^{\dag}_{\mathbf{k}'}
+ a^{\dag}_{\mathbf{p}} 
  a_{\mathbf{p-k+k'}} c_{\mathbf{k}}  c^{\dag}_{\mathbf{k}'} + \ldots
\label{eq:9.72}
\end{eqnarray}

\noindent This expression is represented by two diagrams 3(b) and 3(c). In the diagram 3(b) the
electron line marked $\mathbf{q}$ has been moved to the top of the
 order bar.   In
the diagram 3(c) the product  
$\delta(\mathbf{q} -
\mathbf{p+k})$ and the
integration by $\mathbf{q}$ are represented by  merging or
\emph{pairing}
the 
 incoming electron line carrying momentum
$\mathbf{p-k}$ with the outgoing electron line carrying momentum
$\mathbf{q}$.  This produces the
 \emph{internal electron line} \label{intern} carrying
momentum $\mathbf{p-k}$ between two vertices.

 In the expression (\ref{eq:9.72}), the
 electron operators are in the normal order, however, the photon operators
are not.
The next step is to bring the  photon operators to the normal order

\begin{eqnarray*}
V_1 \underline{V_1} &\propto& 
 a^{\dag}_{\mathbf{q}} a^{\dag}_{\mathbf{p}}a_{\mathbf{p-k}} 
  a_{\mathbf{q+k'}} c^{\dag}_{\mathbf{k}'} c_{\mathbf{k}} 
+   a^{\dag}_{\mathbf{q}} a^{\dag}_{\mathbf{p}}a_{\mathbf{p-k}} 
  a_{\mathbf{q+k'}} \delta(\mathbf{k}'-\mathbf{k}) \\
&+&  a^{\dag}_{\mathbf{p}} 
  a_{\mathbf{p-k+k'}} c^{\dag}_{\mathbf{k}'} c_{\mathbf{k}} 
+  a^{\dag}_{\mathbf{p}} 
  a_{\mathbf{p-k+k'}}\delta(\mathbf{k}'-\mathbf{k})  + \ldots \\
&=&  a^{\dag}_{\mathbf{q}} a^{\dag}_{\mathbf{p}}a_{\mathbf{p-k}} 
  a_{\mathbf{q+k'}} c^{\dag}_{\mathbf{k}'} c_{\mathbf{k}} 
+   a^{\dag}_{\mathbf{q}} a^{\dag}_{\mathbf{p}}a_{\mathbf{p-k}} 
  a_{\mathbf{q+k}}  \\
&+&  a^{\dag}_{\mathbf{p}} 
  a_{\mathbf{p-k+k'}} c^{\dag}_{\mathbf{k}'} c_{\mathbf{k}} 
+  a^{\dag}_{\mathbf{p}} 
  a_{\mathbf{p}}  + \ldots 
\end{eqnarray*}

\noindent  According to equation
(\ref{eq:cc}), the normal ordering of photon operators in 3(b) yields
diagrams 3(d) and 3(e).\footnote{We relabel the external momenta in
Fig. 3(e) for future convenience.}  Diagrams 3(f) and 3(g)
are obtained from 3(c) in a similar manner.

Using diagrams, with some practice,  one can perform 
 calculations of scattering operators  (\ref{eq:8.70}) and (\ref{eq:8.72})
much easier than in the usual algebraic way. 
During these
diagram manipulations we, actually, do not need to keep momentum labels of
lines. 
The
algebraic expression of the result can be easily restored from an
unlabeled diagram by following these steps:

\begin{itemize}
\item[(I)] Assign a distinct momentum label to
each external line, except one, whose momentum is obtained from the
momentum conservation condition. 
\item[(II)] Assign momentum labels to internal lines so
that the momentum conservation law is satisfied at each vertex.  If
there are \emph{loops}, \label{loop} one needs to introduce additional
independent 
\emph{loop
momenta}.\footnote{see diagram 3(g) in which $\mathbf{k}$ is the loop
momentum.} 
\item[(III)]  Read external lines  from top to bottom of the
order bar and write  corresponding particle operators  from left to
right. 
\item[(IV)]  For
each box, write a factor $ i (E_f - E_i)^{-1}$, where
$E_f$ is the sum of energies of particles going out of the box and
$E_i$ is the sum of energies of particles coming into the box. 
\item[(V)]
Write a factor $e^{-iE_Yt}$, where $E_Y$ is the energy
function of the diagram  which is the sum of energies of all outgoing
external lines minus the sum of energies of all incoming external lines. 
\item[(VI)] For each vertex introduce a  factor
$\frac{ e }{\sqrt{(2 \pi )^3 |\mathbf{k}|}}$, where 
$\mathbf{k}$ is the momentum of the photon line attached to this vertex.
\item[(VII)] Integrate the obtained expression by all independent 
external  and loop momenta. 
\end{itemize}

\section { Electron-electron scattering.}
\label{ss:e-e-scatt}

Let us now try to extract  some physical information from the above theory. We will  calculate low order terms in the
perturbation expansion (\ref{eq:8.70}) for the $S$-operator.

\begin{eqnarray}
\Sigma_1(t) &=&   i V_1(t)
\label{eq:9.73}  \\
\Sigma_2(t) &=&   - (V _1(t) \underline{V _1(t)})^{unp}
 - 
(V _1(t)\underline{V _1(t)})^{ph}
  -  (V _1(t) \underline{V _1(t)})^{ren}
                 \label{eq:9.74}  
\end{eqnarray}

\noindent To obtain the corresponding contributions to the $S$-operator we need
to take 
$t$-integrals of these expressions

\begin{eqnarray*}
S = 1 + \underbrace{\Sigma_1(t)} + \underbrace{\Sigma_2(t)} + \ldots
\end{eqnarray*}

\noindent Note that the right hand side of (\ref{eq:9.73}) and the first term on the right
hand side of (\ref{eq:9.74}) are unphys. Their energy
shell is empty,
 so, according to Table 1, they do not
contribute to the $S$-operator. 

 Operator $(V_1 \underline{V}_1)^{ph}$ on the right hand side
of eq. (\ref{eq:9.74}) has two terms corresponding to two types of scattering
processes allowed in the 2nd perturbation order.\footnote{We do not
need to consider unconnected diagrams, like fig. 3(d), because they
describe two or more disjoint scattering processes.} The term of the type $a^{\dag} c^{\dag}ac $
(see, e.g., fig. 3(f)) 
annihilates an electron and a photon in the initial state and
recreates them (with different momenta) in the final state. 
So, this term describes the electron-photon (Compton) scattering.
\label{compton2} In this paper we will focus on the other term which
describes the electron-electron scattering.
Let us consider in more detail the second-order contribution to this process (see fig. 3(e))

\begin{eqnarray*}
S_2[a^{\dag}a^{\dag} aa]
&=&  
2 \pi  \int d \mathbf{p} d \mathbf{q}  d \mathbf{k}
 \delta (\omega_{\mathbf{p-k}} + \omega_{\mathbf{q+k}}
-\omega_{\mathbf{q}}- \omega_{\mathbf{p}}) D_2(\mathbf{p},\mathbf{q}, \mathbf{k})
 a^{\dag}_{\mathbf{p-k}}
a^{\dag}_{\mathbf{q+k}} a_{\mathbf{q}}a_{\mathbf{p}} 
\label{eq:9.76}
\end{eqnarray*}

\noindent  
The coefficient function in (\ref{eq:9.76}) can
be read from the diagram, according to the rules (I) - (VII), 

\begin{eqnarray}
D_2(\mathbf{p},\mathbf{q}, \mathbf{k})
&=&  \frac{i  e^2 }{ (2 \pi )^3 }
 \frac{  1}
{|\mathbf{k}|(|\mathbf{k}| + \omega_{\mathbf{p-k}} -
\omega_{\mathbf{p}})}  
\label{eq:d2}
\end{eqnarray}

\noindent In the non-relativistic approximation ($p, q, k  \ll
mc$),  the coefficient function $D_2(\mathbf{p},\mathbf{q},
\mathbf{k})$ has  singularity $|\mathbf{k}|^{-2}$ 
which is
characteristic for scattering of two  electrons interacting
via repulsive Coulomb potential \label{coulomb1}

\begin{eqnarray*}
\frac{e^2}{4 \pi |\mathbf{r}_1 - \mathbf{r}_2|}.
\end{eqnarray*}

\noindent So, our model theory is quite realistic.

\section { Mass renormalization.}
\label{sc:renormalization}

Next consider the third term on the right hand side of eq. (\ref{eq:9.74}). It is
given by the diagram in fig. 3(g).  According to rules (I) - (VII) this diagram is
represented by the expression

\begin{eqnarray}
(V_1(t) \underline{V_1(t)})^{ren} = \frac{ie^2}{(2\pi)^3} \int  
d\mathbf{p} d\mathbf{k}
 \frac{a^{\dag}_{\mathbf{p}}a_{\mathbf{p}}}
{(\omega_{\mathbf{p}- \mathbf{k}} -\omega_{\mathbf{p}}  + k) k}
\label{eq:9.88}
\end{eqnarray}

\noindent There are  serious problems
with this term. 
First, the loop integral by $\mathbf{k}$ is divergent because the integrand in (\ref{eq:9.88})
has
 asymptotic behavior 
 $ \propto k^{-2}$ at large $k$.
However, even if the integral were convergent, the presence of a renorm
contribution in the operator  $\Sigma(t)$ is unacceptable, because,
according 
to Table 1, the
$t$-integral of any renorm term is infinite. 
 So, the scattering phase in the second order $\underbrace{\Sigma_2}$ is
infinite, which is absurd. Moreover, if we continued
calculations (\ref{eq:9.73}) - (\ref{eq:9.74}) to higher perturbation orders we would find out that
even phys terms in $\Sigma(t)$ become infinite due to divergent loop
integrals,
so the theory with the Hamiltonian (\ref{eq:free-h0}) -
(\ref{eq:9.69}) is seriously flawed.

In a consistent theory we must require that

 \begin{eqnarray}
\Sigma^{ren} = 0
\label{eq:11.42}
\end{eqnarray}

\noindent Therefore, operator $\underbrace{\Sigma}$ must be purely phys

 \begin{eqnarray}
\underbrace{\Sigma} = \underbrace{\Sigma^{ph}}
\label{eq:11.43}
\end{eqnarray}

\noindent It was shown in ref. \cite{GS} that condition (\ref{eq:11.43}) is equivalent to requiring that the
$S$-operator  leaves the vacuum and one-particle
states invariant

\begin{eqnarray}
S | 0 \rangle &=& | 0 \rangle 
\label{eq:11.44} \\
S a^{\dag}_{\mathbf{p}}|0 \rangle &=& a^{\dag}_{\mathbf{p}}|0 \rangle
\label{eq:11.45}\\
S c^{\dag}_{\mathbf{k}}|0 \rangle &=& c^{\dag}_{\mathbf{k}}|0 \rangle
\label{eq:11.46}
\end{eqnarray}

\noindent This is the \emph{mass renormalization condition} of the
traditional renormalization theory.
To satisfy this condition, we must modify  the Hamiltonian (\ref{eq:free-h0}) -
(\ref{eq:9.69})
 by adding certain  unphys $U$ and renorm
$R$ \emph{counterterms} to the interaction operator 
$V_1$.\footnote{It appears that in our model theory the renormalization
is achieved by unphys and renorm counterterms only. In the general
case, e.g., in QED, phys counterterms should be added as well.}
In other words, we are saying that the original Hamiltonian 

 \begin{eqnarray}
H = H_0 + V_1 
\label{eq:org-ham}
\end{eqnarray}

\noindent is not correct, and the modified
 Hamiltonian with renormalization counterterms

 \begin{eqnarray}
H^c &=& H_0 + V^c \nonumber \\ 
&=& H_0 + V_1  +U +R 
\label{eq:ren-ham}
\end{eqnarray}

\noindent better describes interactions between particles.
To comply with eq. (\ref{eq:11.43}), we must choose
the counterterms in such a way that operator 

\begin{eqnarray}
    \Sigma^c(t) &=&
            i V^c(t)
          -  V^c(t) \underline{V^c(t)}
          +   \ldots
\end{eqnarray}

\noindent does not contain renorm terms.
From Table 1, it is clear that renorm terms in $\Sigma^c(t)$ may appear due to
the presence of renorm and unphys terms in $V^c$ and their
products. So,  in order to satisfy eq. (\ref{eq:11.42}), there should
be such a balance between unphysical and renorm terms in $V^c(t)$ that 
all renorm terms in $\Sigma^c(t)$  cancel out in all orders of
the perturbation theory.  The mass
renormalization is achieved by adding renorm counterterms in even
orders: $R_2$, $R_4$, etc. The charge renormalization procedure will
be discussed in section \ref{ss:charge-renorm}. It requires addition of unphys
counterterms in odd orders $U_3(t)$, $U_5(t)$, etc. We take these
considerations into account by writing the general expression for the
Hamiltonian of the renormalized theory\footnote{Of course, we are
looking for a $t$-independent Hamiltonian $H^c$. Although, at
intermediate calculation steps it is convenient to keep all operators
$t$-dependent, as in eq. (\ref{eq:8.67}), in the end we should set $t=0$.}

\begin{eqnarray*}
H^c(t) &=& H_0 + V^c(t) 
\end{eqnarray*}

\noindent where

\begin{eqnarray}
V^c(t) = V_1(t) + R_2 + U_3(t) + R_4 + \ldots
\label{eq:inter}
\end{eqnarray}

\noindent We obtain formulas for $\Sigma^c_i(t)$ ($i = 1,2,3,\ldots$) by inserting
interaction (\ref{eq:inter}) in  (\ref{eq:8.70})
 and collecting terms of equal order.

\begin{eqnarray}
\Sigma^c_1(t) &=&   i V_1(t)
\label{eq:11.47}  \\
\Sigma^c_2(t) &=&   - V _1(t) \underline{V _1(t)}
                +iR_2
\label{eq:11.48} \\
\Sigma^c_3(t) &=&
     - i V_1(t) \underline{V _1(t) \underline{V _1(t)}}
     - R_2 \underline{V _1(t)}
     - V _1(t)  \underline{R_2}
     +iU_3(t)
\label{eq:11.49}  \\
\Sigma^c_4(t) &=&     \sigma_4(t)
     -   U_3(t) \underline{V _1(t)}
     - V_1(t)\underline{U_3(t)} + i R_4
\label{eq:11.50}
\end{eqnarray}

\noindent where we denoted

\begin{eqnarray}
  \sigma_4(t) &= &     V_1(t) \underline{ V_1(t) \underline{V _1(t)
\underline{V _1(t)}}}
     - iV_1(t) \underline{ V_1(t) \underline{R_2}}
\nonumber \\
     &-& i V_1(t) \underline{ R_2 \underline{V _1(t)}}
     - iR_2\underline {V _1(t) \underline{V _1(t)}},\nonumber 
\end{eqnarray}

\noindent Now we go order-by-order and choose counterterms $R_2,
 R_4, \ldots $ so that renorm terms are eliminated from the
left hand sides of eqs (\ref{eq:11.47}) - (\ref{eq:11.50}).  
The first-order term (\ref{eq:11.47}) is
unphys, so there is no need for renormalization in the first order.
To ensure that $\Sigma^c_2(t)  $ does not have a renorm part we choose the
 counterterm

\begin{eqnarray}
  R_{2}  &=& -i(V _1(t) \underline{V _1(t)} )^{ren}
\nonumber 
\end{eqnarray}

\noindent  (see diagram 3(g)). With this choice,  we can rewrite the contributions to the
$S$-operator in the 4 lowest orders

\begin{eqnarray}
\underbrace{\Sigma^c_1(t)} &=&   0 \nonumber \\
\underbrace{\Sigma^c_2(t)} &=&   - \underbrace{(V _1(t) \underline{V _1(t)})^{ph}}
                \nonumber \\
\underbrace{\Sigma^c_3(t)} &=&
     - i \underbrace{V_1(t) (\underline{V _1(t) \underline{V _1(t)}})^{p+u}}
\label{eq:11.51}  \\
\underbrace{\Sigma^c_4(t)} &=&    \underbrace{ \sigma_4(t)}
     -  \underbrace{ U_3(t) \underline{V _1(t)}}
     - \underbrace{V_1(t)\underline{U_3(t)}}  + i  \underbrace{R_4},
\label{eq:11.52}
\end{eqnarray}

\noindent where 

\begin{eqnarray}
 \underbrace{ \sigma_4(t)} &=& 
 \underbrace{ V_1(t) \underline{ V_1(t) (\underline{V _1(t)
\underline{V _1(t)}})^{p+u}}}
+ \underbrace{ V_1(t) \underline{ V_1(t) (\underline{V _1(t)
\underline{V _1(t)}})^{ren}}} \nonumber \\
     &-&i \underbrace{V_1(t) \underline{ R_2 \underline{V _1(t)}}}
     - i \underbrace{R_2\underline {V _1(t) \underline{V _1(t)}}}
     - i \underbrace{V_1(t) \underline{ V_1(t) \underline{R_2}}}
\nonumber \\
&=&
 \underbrace{ V_1(t) \underline{ V_1(t) (\underline{V _1(t)
\underline{V _1(t)}})^{p+u}}}
     -i \underbrace{V_1(t) \underline{ R_2
\underline{V _1(t)}}} \nonumber \\
     &-&i \underbrace{R_2\underline {V _1(t) \underline{V _1(t)}}} 
\label{eq:11.53} 
\end{eqnarray}

\noindent and the superscript $p+u$ denotes the sum of phys and unphys terms.
 The term on the right hand side of (\ref{eq:11.51})
has odd number of particle operators, hence it is free of
renorm parts (which have either zero or two particle operators).
Therefore, no renorm counterterms should be added there. Just as we did in the 2nd order, we can choose a renorm counterterm $R_4$
which simply cancels all renorm terms which may be present in the 
first three terms on the right hand side of (\ref{eq:11.52}).
With the above
choices, the operator $\Sigma^c(t)$  does not contain renorm terms up to
the 4th order,
as required by the mass renormalization condition,\footnote{Our analysis demonstrates that that mass
renormalization is always  necessary when interaction contains unphys terms,
like trilinear operators in eq. (\ref{eq:9.69}) and in more realistic
theories, such as QED, QCD, and Standard Model. Products or commutators of such terms in 
(\ref{eq:8.70}) or (\ref{eq:8.72}) give rise to renorm terms in the
$S$-operator
which should be compensated by adding renorm terms to the Hamiltonian.
The operator structure of these terms is the same as in $H_0$  
(\ref{eq:free-h0}). This
results in the difference of masses of bare and dressed particles, as
discussed at the end of section \ref{ss:charge-renorm}.} 
and  the expression for
$\underbrace{\Sigma^c_4}$ simplifies

\begin{eqnarray}
  \underbrace{\Sigma^c_4(t) }
 &=&  - \underbrace{((\underline{V_1(t)}  V_1(t))^{p+u} (\underline{V _1(t)
\underline{V _1(t)}})^{p+u})^{ph}} \nonumber \\
     &+& \underbrace{(\underline{V_1(t)}  
(V_1(t) \underline{V_1(t)})^{ren} \underline{V _1(t)})^{ph}}
-  \underbrace{ (U_3(t) \underline{V _1(t)})^{ph}}
     - \underbrace{(V_1(t)\underline{U_3(t)})^{ph}}
\label{eq:11.55}  
\end{eqnarray}

Using  the diagram
technique we find that the $a^{\dag}a^{\dag}aa$ part of the first two
terms on the right hand side of eq. (\ref{eq:11.55})  is represented by 7 diagrams shown in Fig.
4.   From diagram rules (I) - (VII), we obtain the contribution of diagrams 4(a) - 4(b) to
the coefficient function $D$ for the electron-electron scattering on the
energy shell.

\begin{eqnarray}
D_4^{(a - b)}(\mathbf{p}, \mathbf{q}, \mathbf{k}) = 
- \frac{i  e^4  }{(2\pi)^6 
(\omega_{\mathbf{p-k}} -\omega_{\mathbf{p}} + k)k}
\int \frac{d\mathbf{h}}{h}
(\frac{1}{BC} + \frac{1}{EF})
\label{eq:11.58}
\end{eqnarray}

\noindent where $B = \omega_{\mathbf{p-h}}
-\omega_{\mathbf{p}}  + h$, $C = \omega_{\mathbf{p-k}}
-\omega_{\mathbf{p-h-k}}  - h$,  $E = \omega_{\mathbf{q-h}}
-\omega_{\mathbf{q}}  + h$, and $F = \omega_{\mathbf{q+k}}
-\omega_{\mathbf{q+k-h}} - h $. Unfortunately, this contribution  is infinite (at
large values of $\mathbf{h}$ the integrand behaves as $h^{-3}$).
 Thus we conclude that
 the mass renormalization procedure described above has not
removed all divergences. To solve this problem we need to perform the
second renormalization step known as the charge renormalization
procedure. This step is explained in the next section.

\begin{figure}
\epsfig {file=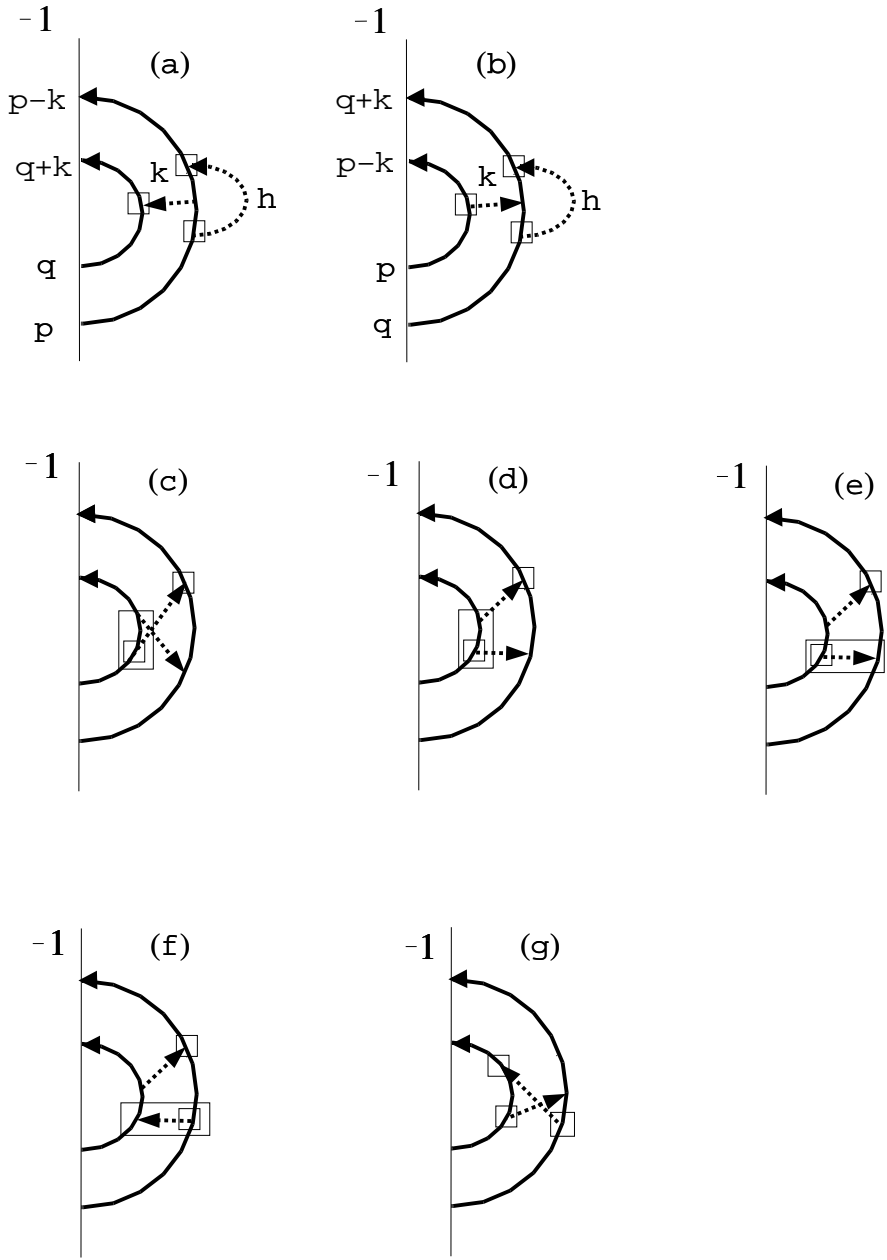}
\caption{Contributions with the operator  structure
$a^{\dag} a^{\dag} aa$ 
to the first two terms on the right hand side of eq. (\ref{eq:11.55}).}
\end{figure}

\section{ Charge renormalization}
\label{ss:charge-renorm}

First note that  the divergent terms (\ref{eq:11.58})  have a singularity $k^{-2}$ at
$k \to 0$.
 As we know from section \ref{ss:e-e-scatt}, such a singularity is responsible
for the low-energy electron-electron scattering at large distances. From
classical physics we also know that long-distance interactions between
charged particles depend on $e^2$ (in our language,
they are of the second perturbation order) and
 they are accurately described by the 2nd order term (\ref{eq:d2}). 
Non-zero terms $D_4^{(a - b)}$ mean that the charge of the electron is
modified by the interaction. Actually, this modification is infinite,
but even finite values of the terms like (\ref{eq:11.58}) are inconsistent with the
classical limit.
 So, we will
postulate that in orders higher that 2nd, singular coefficient
functions  like (\ref{eq:11.58}) should not be present at all, whether they
are infinite or finite.
This is the \emph{charge renormalization condition}. To eliminate the
divergent  contribution (\ref{eq:11.58}) we can choose  unphys
3rd order counterterms $U_3$.
Two diagrams describing these counterterms are shown in Fig. 5(a) and 5(b). 
As expected, they are infinite, and expressions  

\begin{eqnarray*}
-   U_3(t) \underline{V _1(t)}
     - V_1(t)\underline{U_3(t)}
\end{eqnarray*}

\noindent in (\ref{eq:11.52})  exactly  
cancel unwanted infinite diagrams 4(a)
and 4(b)   on the energy shell.

\begin{figure}
\epsfig {file=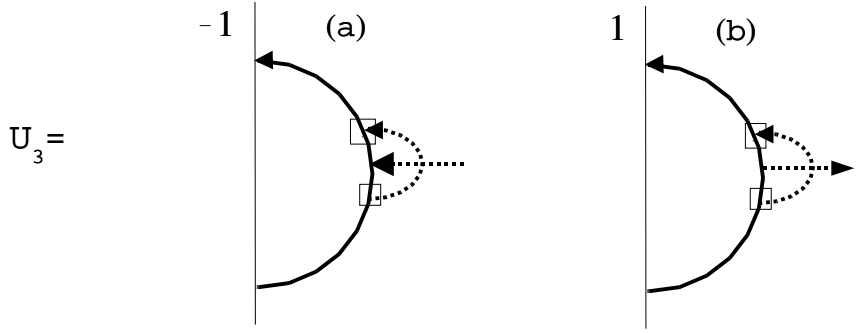}
\caption{ Charge renormalization counterterms $U_3(t)$. }
\end{figure}

Surviving diagrams 4(c) - 4(g)
are the 4th order
\emph{radiative corrections} \label{rad-corr} to the electron-electron
scattering. On the
energy shell they yield the following contribution to the 
coefficient function of $\Sigma(t)$

\begin{eqnarray}
D_4^{(c-g)}(\mathbf{p}, \mathbf{q}, \mathbf{k}) 
&=& - \frac{i  e^4   }{(2\pi)^6 }
\int \frac{ d \mathbf{h}}{h|  \mathbf{h}  + \mathbf{k} |}
\frac{1}{A}(\frac{1}{BC}
+ \frac{1}{DC}+ \frac{1}{EF}+ \frac{1}{DG}+ \frac{1}{EG})
\label{eq:11.59}
\end{eqnarray}

\noindent where

\begin{eqnarray*}
A&=&\omega_{\mathbf{q-h}}-\omega_{\mathbf{q}}  +h \\
B&=&\omega_{\mathbf{p-k-h}} -\omega_{\mathbf{p}} +|\mathbf{h+k}|\\
C&=&\omega_{\mathbf{q-h}} +\omega_{\mathbf{p-k}}+|\mathbf{h+k}|-
\omega_{\mathbf{q}} -\omega_{\mathbf{p}}\\
D&=&\omega_{\mathbf{p+h}} +\omega_{\mathbf{q-k}} -
\omega_{\mathbf{q}} -\omega_{\mathbf{p}} \\
E&=&\omega_{\mathbf{q+k}} -\omega_{\mathbf{q}}+|\mathbf{h+k}| + h\\
F&=&\omega_{\mathbf{q+k}} +\omega_{\mathbf{p-h-k}} +h -
\omega_{\mathbf{q}} -\omega_{\mathbf{p}}\\
G&=&\omega_{\mathbf{q+k}} +\omega_{\mathbf{p+h}} +|\mathbf{h+s}| -
\omega_{\mathbf{q}} -\omega_{\mathbf{p}}
\end{eqnarray*}

\noindent The integrand in (\ref{eq:11.59}) is
proportional to $h^{-5}$,
and the integral is convergent at large values of the loop momentum
$\mathbf{h}$.
Therefore
$\Sigma_4^c(t)$ is finite on the energy 
shell. So, in the
renormalized theory 
the electron-electron scattering is represented by a finite
 $S$-operator which up to the 4th order is described by the 
coefficient function 

\begin{eqnarray*}
D = D_2 + D_4 + \ldots
\end{eqnarray*}

\noindent where $D_2$ is given by eq. (\ref{eq:d2}) and $D_4$ is given by eq. (\ref{eq:11.59}).
\label{charge-end}

The renormalization technique presented above has not been applied to
realistic theories, such as QED, yet. However, the close analogy
between renormalization issues encountered in our toy model and in QED
allows us to speculate that similar renormalization steps
can be repeated in all perturbation orders in QED, so that finite and
accurate values for various scattering amplitudes can be obtained.
Note that in our approach the
counterterms in the Hamiltonian (\ref{eq:ren-ham}) have the same operator structure as
the terms in the original Hamiltonian (\ref{eq:org-ham}). In relativistic
renormalizable theories, like QED, even stronger statements can be
made:
the mass renormalization
counterterms  differ from the one-particle energy terms in $H_0$ only by a
constant (but infinite) factor,\footnote{From this point of view the
mass renormalization was discussed in
ref. \cite{Kruger}.} and the charge renormalization
counterterms are equal to the interaction term $V$ in the original
Hamiltonian
 multiplied by a constant
infinite factor. This conclusion is true
in all perturbation orders. Thus, if the original Hamiltonian depends on finite masses
$m_1$, $m_2, \ldots$ and coupling constants (charges) $e_1$, $e_2, \ldots$

\begin{eqnarray}
H(m_1, m_2, \ldots; e_1, e_2, \ldots)
\label{eq:11.60}
\end{eqnarray}

\noindent   Then
the addition of renormalization
 counterterms is equivalent to simply modifying (making them infinite)
the values of 
masses and charges.
The Hamiltonian $H^c$ after renormalization has the same functional
form as $H$ where parameters $m_1$, $m_2, \ldots$ and  
$e_1$, $e_2, \ldots$ substituted by renormalized (infinite) values
$\tilde{m_1}$, $\tilde{m_2}, \ldots$ and  
$\tilde{e}_1$, $\tilde{e}_2, \ldots$

\begin{eqnarray}
H^c &=& H(\tilde{m}_1, \tilde{m}_2, \ldots; \tilde{e}_1, \tilde{e}_2, \ldots)
\end{eqnarray}

The traditional
approach offers the following physical interpretation of these
results. The creation and annihilation operators present in the theory
describe so-called bare particles having infinite masses and
charges. The interaction between bare particles drastically change
their properties. For example, bare electrons constantly emit and
re-absorb virtual photons and electron-positron pairs. So, bare
electrons
are surrounded by a coat of virtual particles. The mass of this coat
is infinite, thus compensating the infinite mass of the bare particle.
The virtual particles in the coat shield the true
(infinite) charge of the bare electron. The resulting mass and
charge of the dressed particle is finite and exactly equal to the
parameters measured in experiment. In experiment, we never see the
bare particles and their virtual coats, we only see the dressed
particles. 

The Hamiltonian $H^c$ is formally infinite, but these infinities
cancel when the $S$-operator is calculated by formula (\ref{eq:8.68}), and accurate
results are obtained for scattering amplitudes and energies of bound
states. However, the infinities do not cancel when
one tries to find the eigenvectors of the Hamiltonian $H^c$ or
calculate the time evolution operator $\exp(iH^ct)$. Direct
application of this formula would lead to unphysical results. For
example, if we  calculated the time evolution of the simplest
one-electron state we would obtain that this states dissociates into 
a complex linear combination of states over time

\begin{eqnarray}
e^{iH^ct} a^{\dag} |0> &=& (1 + iH^ct + \ldots)  a^{\dag} |0>
\nonumber \\
& \propto & (1 + it(a^{\dag}a + c^{\dag}c + a^{\dag}c^{\dag}a
+ a^{\dag}a c) + \ldots)  a^{\dag} |0> \nonumber \\
& \propto &   a^{\dag} |0> + a^{\dag}c^{\dag} |0> +
a^{\dag}c^{\dag}c^{\dag} |0> + \ldots
\label{eq:ihct}
\end{eqnarray}

\noindent Moreover the coefficients multiplying the terms in (\ref{eq:ihct}) are
given by divergent integrals.
Therefore, with the infinite Hamiltonian $H^c$ the basic formulas of quantum
theory (\ref{eq:1}) and (\ref{eq:8.60}) become useless for practical
calculations.
 However,
in most cases this is not a cause of trouble. As discussed in
section \ref{sc:bound-states}, most experiments in high energy physics are concerned
either with bound state energies or with scattering cross-sections. To
calculate these properties, the knowledge of the $S$-operator is
sufficient and the renormalized theory works fine. For approximate description of
the time evolution in low energy, e.g., atomic, systems, one can use 
heuristic approaches, such as those based on the Dirac-Fock
Hamiltonian.

\section {Construction of the dressed particle Hamiltonian}
\label{sc:constr}

We see two fundamental problems with the traditional renormalization
theory presented above. First, from the practical point of view, 
the absence of a well-defined finite Hamiltonian does not allow one to
study the time evolution of interacting states. One can expect that
with advancement of experimental tools, the time-dependent information
from the region of interaction will soon become available. However, 
accurate analysis of this information is not possible with 
ill-defined Hamiltonians of renormalized
quantum field theories.\footnote{ The way to remove the ultraviolet
divergences from interaction operators, or to ``renormalize'' the
Hamiltonian, was suggested by G\l azek and Wilson in their similarity
renormalization approach \cite{Glazek}. However, their Hamiltonians
still have unphys terms \cite{Glazek2}, and result in ``instability'' of vacuum and
one-particle states similar to that shown in eq. (\ref{eq:ihct}).} 
Second, from the theoretical point of view, the traditional
approach operates with notions of bare and virtual particles that
are, in principle, non-observable. It is desirable to have a theory
formulated directly in terms of dressed particles and
their interactions without mentioning the non-observable bare and
virtual particles
at all. This idea was first realized in the dressed particle
formalism by Greenberg and Schweber \cite{GS}. 

In the rest of this paper we will discuss the 
 RQD approach \cite{infinities, Stefanovich_Mink,  mybook} which combines the similarity
renormalization and dressed particle ideas. RQD is capable  to fix the
both problems of the traditional renormalization formalism mentioned above.
Let us first focus on the dressing part of RQD.
 There are two approaches to dressing with different interpretations
but equivalent physical results. One approach \cite{shebeko} tries
to find 
an explicit
unitary (dressing) transformation connecting creation and annihilation
operators of bare particles ($a^{\dag}, a, c^{\dag}$, and $c$) with
creation and annihilation operators of dressed particles ($A^{\dag},
A, C^{\dag}$, and $C$)
and find a function $f$ which  expresses the Hamiltonian $H^c$ of the 
renormalized theory through
dressed operators

\begin{eqnarray}
H^c = f(A^{\dag},
A, C^{\dag}, C)
\end{eqnarray} 

\noindent Another approach \cite{infinities, Stefanovich_Mink,  mybook}  is to pretend that operators 
$a^{\dag}, a, c^{\dag}$, and $c$ already describe the dressed
particles and try to find a new finite Hamiltonian $H^d$ whose functional 
dependence on bare
particle operators is given by the same function $f$ 
which expresses the dependence of $H^c$ on the dressed
operators

\begin{eqnarray}
H^d = f(a^{\dag},
a, c^{\dag}, c)
\end{eqnarray}

\noindent We will stick to the second approach in this paper. There
are two ways to proceed. First, we can try to find $H^d$ by applying a
unitary dressing transformation to $H^c$. This way was described
in ref. \cite{infinities}. In this paper we are choosing another
route: we will simply fit
the Hamiltonian $H^d$ to the finite scattering operator $S^c$ known from the
renormalized theory.

There are three requirements that we want to satisfy when looking for the
``dressed
particle'' Hamiltonian $H^d = H_0 + V^d = H_0 + V^d_2 + V^d_3 + V^d_4 + \ldots$.

\begin{itemize}
\item[(A)] $H^d$ is \emph{scattering-equivalent} to $H^c$.
\item[(B)] $V^d$ is \emph{finite}. 
\item[(C)] $V^d$ is \emph{phys}. 
\end{itemize}

The condition (A) is understandable as we know that the $S$-operator
of  renormalized theories (QED, Standard Model) agrees well with
experiments, and we would like to preserve this agreement in a theory with the new
Hamiltonian $H^d$.
Using eqs. (\ref{eq:8.71}) and (\ref{eq:8.72}) we can write

\begin{eqnarray}
  -i \log  S^c &=&   \underbrace{F^c_2(t)} + \underbrace{F^c_3(t)}  +
\underbrace{F^c_4(t)} + \ldots  \nonumber \\
  -i \log    S^d    &=&  \underbrace {V^d_2(t)}
    + \underbrace {V^d_3(t)} +  
\underbrace {V^d_4(t)} - \frac{i}{2 } \underbrace
{[\underline {V^d_2(t)},V^d_2(t)]}  \ldots  \nonumber
\end{eqnarray}

\noindent Then we see that condition (A) implies  the following
infinite set of
relations between $V_i^d(t)$ and $F_i^c(t)$ on the energy
shell\footnote{The use of the Magnus expansion for deriving $V_i^d(t)$ is preferable to
the old-fashioned or Dyson's formulas, because the right hand sides of
eqs. (\ref{eq:12.4}) - (\ref{eq:12.7}) are expressed through commutators, so they are manifestly
Hermitian. In addition, this approach automatically generates cluster
separable interactions $V^d_i(t)$ (see ref. \cite{infinities}).}

\begin{eqnarray}
\underbrace{V_2^d(t)} &=&  \underbrace{F_2^c(t)}
\label{eq:12.4} \\
\underbrace{V_3^d(t)} &=& \underbrace{F_3^c(t)}
\label{eq:12.5} \\
\underbrace{V_4^d(t)} &=& \underbrace{F_4^c(t)} + \frac{i}{2} \underbrace
{[\underline {V^d_2(t)},V^d_2(t)]} 
\label{eq:12.6}\\
\underbrace{V_i^d(t)} &=& \underbrace{F_i^c(t)} + \underbrace{Q_i(t)},
\mbox{        } i > 4
\label{eq:12.7}
\end{eqnarray}

\noindent where  $Q_i(t)$  denotes a  sum of multiple commutators of $V_j^d(t)$ from
lower orders ($2 \leq j \leq i-2$) with $t$-integrations.
We have expressions for operators $\Sigma_2^c(t)$ and $\Sigma_4^c(t)$
in the renormalized theory (eqs. (\ref{eq:d2}) and (\ref{eq:11.59})). Clearly, we can perform
similar calculations for operators $F_i^c(t)$ whose values on the
energy shell are present on the right hand sides of eqs. 
(\ref{eq:12.4}) - (\ref{eq:12.7}).  Alternatively, we can express
$F^c(t)$ through $\Sigma^c(t)$ using relationship (\ref{eq:8.74}).

Operators $F_i^c(t)$ are, of course,  finite. 
This immediately
implies that $V^d_2(t)$ and $V^d_3(t)$ are finite on the energy shell.
Moreover, from eqs. (\ref{eq:8.74}) and (\ref{eq:11.43}) it follows that $\underbrace{F_2^c(t)}$ and 
$\underbrace{F_3^c(t)}$
 are phys on the energy shell. Therefore, $V_2^d(t)$ and $V_3^d(t)$  can be also
chosen phys on the energy shell. This proves that conditions (B) and
(C) are satisfied on the energy shell up to the 3rd order. 
However, eqs. (\ref{eq:12.4}) - (\ref{eq:12.7}) tell us nothing about the behavior of $V^d_2(t)$
and $V^d_3(t)$ off the energy shell.\footnote{Note that in
applications we are dealing primarily with interactions near the
energy shell, because in most processes, excluding very short virtual
events, the total energy stays unchanged at all times.} The same is true for
interactions in higher orders: the interaction operators $V^d_i(t)$ off the
energy shell remain undetermined by our condition (A). This freedom in
choosing interactions simply means that different choices of the
coefficient functions of $V^d_i(t)$ off the energy shell result in
scattering equivalent Hamiltonians. Since our experimental knowledge
about interactions in high energy physics is limited to their effects
on scattering, we are allowed to freely choose the behavior of
$V^d_i(t)$ off the energy shell without any danger to get into
contradiction with experiment. This freedom is exactly what is needed
to satisfy conditions (B) and (C) in all orders. As we will see
shortly, it is important to choose the behavior of
$V_i^d(t)$  such that their coefficient
functions fall off rapidly when the arguments move away from the energy
shell.\footnote{In the representation where $H_0$ is diagonal, this condition is
equivalent to bringing the matrix of the Hamiltonian to the
band-diagonal form, which is the central idea of the similarity
renormalization method \cite{Glazek}.} For example, we can choose the coefficient functions being
proportional to $\zeta_i = \exp(- \gamma E^2)$ where $\gamma$ is a positive
constant and $E$ is the energy function.\footnote{Note that we should
be careful not to set $\gamma$ to infinity. In this case, the
coefficient functions of interaction operators $V_i^d$ become
non-differentiable. This means that interaction is not separable
\cite{book}, 
and the scattering theory formalism from section \ref{sc:bound-states}
is no longer applicable.}
 Then the
electron-electron interaction in the second order takes the form

\begin{eqnarray}
V^d_2(t)
&=&  - \frac{i}{2 } [\underline{V_1^c(t)}, V_1^c(t)]^{ph} \circ
\zeta_2 \nonumber \\
&=& \frac{  e^2 }{ (2 \pi )^3 }
  \int d \mathbf{p} d \mathbf{q}  d \mathbf{k}
 \frac{  e^{- \gamma E^2}e^{-it E}}
{|\mathbf{k}|(|\mathbf{k}| + \omega_{\mathbf{p-k}} - \omega_{\mathbf{p}})}  
 a^{\dag}_{\mathbf{p-k}}
a^{\dag}_{\mathbf{q+k}} a_{\mathbf{q}}a_{\mathbf{p}} 
\label{eq:el-el}
\end{eqnarray}

\noindent where $E = \omega_{\mathbf{p-k}} + \omega_{\mathbf{q+k}}
-\omega_{\mathbf{q}}- \omega_{\mathbf{p}}$ is the energy function.

There are no terms of the type $a^{\dag}a^{\dag}aa$ in the 3rd order
term $F_3^c(t)$. Therefore the lowest (4th) order radiative correction
to the electron-electron interaction (\ref{eq:el-el}) should be obtained from
eq. (\ref{eq:12.6}). As discussed above, the value of $F_4^c(t)$ on the energy
shell is finite.
 Let us now consider  the  term 

\begin{eqnarray}
\frac{i}{2} \underbrace{[\underline
{V^d_2(t)},V^d_2(t)]}
\label{eq:12.10}
\end{eqnarray}

\noindent   on the right hand side of eq. (\ref{eq:12.6}). First, we note that,
according to Table 1,
this
commutator is phys. Second, as we agreed above, the
coefficient functions of $V^d_2(t)$  fall off rapidly outside the energy
shell. Without this condition, the loop integrals encountered in
calculations of (\ref{eq:12.10}) could be divergent. However, in our case these
integrals are convergent: when the loop integration momentum goes to
infinity, the $V^d_2(t)$ factors in the integrand go away from the
energy shell, i.e., rapidly fall off. This guarantees that
interaction $V^d_4(t)$ is phys and finite on the energy shell. Its coefficient function
off the
energy shell should be again chosen to decay rapidly  to
ensure the convergence of loop integrals in higher order  operators
$Q_i(t)$,
where $V^d_4(t)$ may contribute. These arguments can be repeated in higher orders, which proves that
the dressed particle Hamiltonian $H^d$ is free of ultraviolet divergences.

The
contribution  $(F^c_4)^{ph}$ on the right hand side of eq. (\ref{eq:12.6}) is
well-defined near 
the energy shell,
but this is not true for the contribution  
$\frac{i}{2} [\underline {V^d_2},V^d_2] $.
This
commutator
 depends on the behavior of $V^d_2$
everywhere in the  momentum space. So, it substantially depends on our
choice of $\zeta_2$ off the energy shell. 
There is a great freedom in this choice
which is reflected in the uncertainty of $V_4^d$ even on the energy shell. 
 It is interesting to note that
although the off-shell behavior 	of the 2nd order interaction and
the on-shell behavior of the 4th order interaction cannot be separately  determined
in our theory, they are connected to each other in such a way that the
ambiguity of the interaction does not affect the $S$-matrix for the
electron-electron scattering.

The above construction does not
allow us to
obtain full information about $V^d$: The off-shell behavior of
interactions is rather arbitrary, and the on-shell behavior
can be determined only for lowest order terms.
However, this uncertainty is perfectly understandable: It simply reflects the one-to-many correspondence
between the $S$-operator and Hamiltonians (see section \ref{sc:bound-states}).
It means that there is a class of finite phys
interactions $ \{V^d\} $ all of which satisfy
our requirements (A) - (C) and can be used for $S$-matrix calculations without
encountering divergent integrals. 
 In order to find the unique Hamiltonian
correctly describing the dynamics of particles, the theoretical
predictions should be compared with time-resolved experimental
data. However, our model theory is not sufficiently accurate to be
comparable with experiment, and further efforts
in this direction require building the dressed particle formulation of
the full-blown quantum electrodynamics \cite{infinities, mybook}.

\begin{table}[h]
\caption{Examples of interaction terms in the dressed particle
Hamiltonian (\ref{eq:dress}).  Bold
numbers in the third  column  indicate perturbation orders  in which
the operators can be unambiguously obtained near the energy shell
  as discussed in section
\ref{sc:constr}.}
\begin{tabular*}{\textwidth}{@{\extracolsep{\fill}}lll}
\hline

 Operator     &Physical meaning        &    Perturbation     \cr
                 &               &   Orders      \cr
\hline
   & \bf{Elastic potentials} & \cr
$a^{\dag}a^{\dag}aa$ &electron-electron  & $\mathbf{2},4,6, \ldots$ \cr
$a^{\dag}c^{\dag}ac$  &electron-photon (Compton)  &  $\mathbf{2},4,6,\ldots$ \cr
$a^{\dag}a^{\dag}a^{\dag}aaa$ & 3-electron potential & $4,6,\ldots$
\cr
 & \bf{Inelastic potentials} & \cr
$a^{\dag}a^{\dag}c^{\dag}aa$ & bremsstrahlung in electron-electron collisions &
$\mathbf{3},5, \ldots $\cr
$a^{\dag}a^{\dag}aac$ & photon absorption in electron-electron collisions &
$\mathbf{3},5, \ldots $ \cr
\hline
\end{tabular*}
\end{table}

In
contrast to the original Hamiltonian (\ref{eq:org-ham}), there seems to be no way to
write the dressed particle Hamiltonian $H^d$ in a closed form. From the derivation outlined above it is
clear 
that in higher perturbation orders there are more and more
terms  with increasing complexity in the interaction operator $V^d$.

\begin{eqnarray}
H^d &=& H_0 + V^d \nonumber \\
&=&  H_0 + a^{\dag}a^{\dag}aa + a^{\dag}c^{\dag}ac +
a^{\dag}a^{\dag}a^{\dag}aaa 
+ a^{\dag}a^{\dag}aac + a^{\dag}a^{\dag}c^{\dag}aa + \ldots
\label{eq:dress}
\end{eqnarray} 

\noindent Some of
them are shown in Table 2.  However, all these
high order terms directly reflect real interactions and processes
observable in nature.  For example, the term 
$a^{\dag}a^{\dag}c^{\dag}aa$ (\emph{bremsstrahlung})
\label{bremsstr} describes
creation 
of a photon in electron-electron  collisions. In the  language of classical
electrodynamics, this can be
interpreted as radiation due to the acceleration  during
interaction of charged particles and is often referred to as the 
\emph{radiation reaction} \label{rad-res} force. 
The Hermitian-conjugated term $a^{\dag}a^{\dag}aac$
describes absorption of a photon by a colliding pair of charged 
particles.

Note that in contrast to the traditional renormalization approach in
which the mass renormalization condition (\ref{eq:11.42}) was ensured 
by maintaining a balance between unphys and renorm terms in the
interaction, in the RQD Hamiltonian these terms are absent
altogether. The interaction operator $V^d$ is purely phys which
guarantees that scattering operators $\Sigma^d$ and $F^d$ are phys
too. The phys character of the interaction $V^d$ means that it
yields zero when acting on the vacuum and one-particle states. 
Interaction $V^d$ acts only when there are two or more dressed 
particles present. This is equivalent to saying   
that  self-interaction effects are not present in the dressed particle
approach. Naturally, there should be no dressing in
our theory, because we are working with particles which are already fully
dressed.

\section* {Conclusions}

In this paper, we considered a model theory which provides a
simplified description of interactions between electrons and photons,
similar to exact interactions in QED. We found that in order to get
sensible results for the $S$-matrix, the theory should be
renormalized, just as QED, by adding infinite counterterms to the
Hamiltonian. 
We also demonstrated that the renormalized version of the  theory
 can be reformulated entirely in terms of dressed
particles and their interactions without affecting predictions of the
theory about scattering cross-sections or bound state energies. 
The dressed particle Hamiltonian
$H^d$ is finite, thus, the
 ultraviolet divergences are not present anymore. 
There are no bare and virtual particles in the dressed particle
approach. This simplifies substantially the physical interpretation of
the theory. From the operator form of the dressed particle Hamiltonian
(\ref{eq:dress})
it is clear that dressed particles interact with each
other via instantaneous forces that generally do not conserve the
number of photons.

\section* {Acknowledgement}

I am thankful to Drs. Igor Khavkine, Arnold Neumaier, and 
Dan Solomon for helpful comments, discussions,
and criticism and to Dr. Arnold Neumaier for drawing my attention to 
ref. \cite{Glazek} and related works.

\end{document}